\documentclass[times,doublespace]{wcmauth}
\usepackage{CJK}
\usepackage{indentfirst}
\usepackage{color}
\usepackage[square, comma, sort&compress, numbers]{natbib}
\begin{document}
\runningheads{Xiaohu Ge \emph{et~al.}}{5G Multimedia Massive MIMO Communications Systems}

\articletype{RESEARCH ARTICLE}

\title{5G Multimedia Massive MIMO Communications Systems}

\author{Xiaohu Ge\affil{1}, Haichao Wang\affil{1}, Ran Zi\affil{1}, Qiang Li\affil{1} and Qiang Ni\affil{2}}

\address{\affilnum{1}School of Electronic Information and Communications, Huazhong University of Science \& Technology, Wuhan, China.\\
\affilnum{2}School of Computing and Communications, Lancaster University, Lancaster LA1 4WA, UK.}

\corraddr{Dr. Qiang Li, School of Electronic Information and Communications, Huazhong University of Science and Technology, Wuhan 430074, Hubei, P. R.China. E-mail: qli\_patrick@mail.hust.edu.cn \\
Part of this work appeared in the IEEE IWCMC 2015 \cite{Wang15}, which was granted the best paper award. }


\begin{abstract}
In the Fifth generation (5G) wireless communication systems, a majority of the traffic demands is contributed by various multimedia applications. To support the future 5G multimedia communication systems, the massive multiple-input multiple-output (MIMO) technique is recognized as a key enabler due to its high spectral efficiency. The massive antennas and radio frequency (RF) chains not only improve the implementation cost of 5G wireless communication systems but also result in an intense mutual coupling effect among antennas because of the limited space for deploying antennas. To reduce the cost, an optimal equivalent precoding matrix with the minimum number of RF chains is proposed for 5G multimedia massive MIMO communication systems considering the mutual coupling effect. Moreover, an upper bound of the effective capacity is derived for 5G multimedia massive MIMO communication systems. Two antenna receive diversity gain models are built and analyzed. The impacts of the antenna spacing, the number of antennas, the quality of service (QoS) statistical exponent, and the number of independent incident directions on the effective capacity of 5G multimedia massive MIMO communication systems are analyzed. Comparing with the conventional zero-forcing precoding matrix, simulation results demonstrate that the proposed optimal equivalent precoding matrix can achieve a higher achievable rate for 5G multimedia massive MIMO communication systems.
\end{abstract}


\maketitle

\section{Introduction}
\label{sec1}

As various wireless multimedia applications are getting more and more popular, the demand for wireless traffic is increasing rapidly, and the massive multi-input-multi- output (MIMO) technology has been proposed as a key technology for the next generation (5G) wireless communication systems \cite{Larsson13,Ge14,Nadembega14}. Recently, a number of excellent studies have validated that massive MIMO systems are specialized in improving the wireless communication capacity vastly in cellular networks\cite{Rusek13}. Apparently the huge antenna arrays have to be deployed compactly because enough space are not available at not only base stations (BSs) but also mobile terminals, therefor the interaction of mutual coupling among antennas gets so strong that it can't be ignored in massive MIMO systems\cite{xu10}. Also, the realistic channel capacity which is subject to the quality of service (QoS) in multimedia wireless communication systems and the Shannon capacity are not the same thing \cite{Fernandez09,Niyato10}. So, exploring a new precoding solution for the 5G massive MIMO multimedia communication systems is necessary.

A lot of studies have achieved great achievements about mutual coupling among multiple antennas on many topics such as antenna propagation, signal processing and antenna arrays \cite{Andersen76,Gupta83,B.Clerckx07,Steyskal90}. Utilizing the real measurement data, the authors of \cite{Andersen76} have made a comparison on the antenna array performance between the systems considering the mutual coupling and the systems not. It has been proved that mutual coupling has a great influence on the performance of antenna arrays for not only small but also large inter-antenna spacing, because that in order to contain the changes in all the anticipant vectors, the steering vectors of the antenna arrays should be adjusted not only in amplitude but also in phase\cite{Gupta83}. Clerckx \emph{et. al.} studied how the mutual coupling influenced a simple multi-antenna communication system performance \cite{B.Clerckx07}. In order to recover the signals received by separate antennas without mutual coupling, the authors of \cite{Steyskal90} have invented a new technique to make a compensation for mutual coupling in small antenna arrays.

At practical wireless communication transmission terminals, each data stream is first passed through the baseband precoding to radio frequency (RF) chains and then is transmitted to antennas by the RF chains precoding. For MIMO wireless systems, the precoding technologies are focused on the baseband precoding, i.e., the first order precoding, and each RF chain corresponds to an antenna. Utilizing the phase matrix between RF chains and antennas, the joint precoding of baseband and RF chains was proposed for massive MIMO systems with limited RF chains \cite{Liu14}. However, it is still a great challenge to reduce the number of RF chains for saving the cost of massive MIMO wireless communication systems.

Lots of excellent studies in the field of wireless multimedia communication have emerged \cite{Wu03,Taleb08,Tang071,Taleb07,Tang072}. In order to evaluate the QoS of wireless multimedia networks, the authors in \cite{Wu03,Taleb08} created a constrained model of statistical QoS to study the transmission characteristics of data queues. In \cite{Tang071,Taleb07}, the authors referred to the effective capacity of the block fading channel model and proposed a rate and power adaption scheme in which the power is driven by QoS. And in \cite{Tang072}, the authors further combined the effective capacity with information theory and developed some rate adaptation and QoS-driven power schemes which were suitable for the systems of multiplexing and diversity. Also they concluded that stringent QoS and high throughput can be achieved by the multi-channel communication systems simultaneously according to their simulation results. However, rare efforts has been made to study the effective capacity of massive MIMO multimedia wireless communication systems which consider the QoS constraint and mutual coupling effect.

Motivated by the above gaps, we propose an optimal equivalent precoding matrix to reduce the cost of RF chains in 5G massive MIMO multimedia communication systems and derive the upper bound of effective capacity with QoS constraints. The main contributions of this paper are listed as follows.

\begin{enumerate}
\item We define the receive diversity gain to analyze how the mutual coupling influence the performance of the rectangular antenna arrays in the massive MIMO wireless communication systems.
\item An optimal equivalent precoding matrix is proposed to reduce the cost of RF chains and satisfy the multimedia data requirements for 5G massive MIMO multimedia communication systems.
\item We refer to the QoS statistical exponent constraint and the mutual coupling effect, then derive the upper bound of effective capacity for 5G massive MIMO multimedia communication systems.
\item Based on numerical results, the proposed optimal equivalent precoding matrix is compared with the conventional zero-forcing (ZF) precoding matrix in 5G massive MIMO multimedia communication systems.
\end{enumerate}

The rest of this paper is summarized as follows. In Section 2, a system model in which there is a 2D antenna array is described for massive MIMO wireless communications. In Section 3, the effect of mutual coupling on the massive MIMO wireless systems is evaluated by the receive diversity gain. Moreover, an optimal equivalent precoding matrix is proposed to reduce the cost of RF chains and satisfy the multimedia data requirements for 5G massive MIMO multimedia communication systems. Furthermore, the upper bound of effective capacity is derived for 5G massive MIMO multimedia communication systems. Numerical simulations and analysis are presented in Section 4. Finally, Section 5 summarizes the paper.

\section{System model}

\begin{figure}
  \centering
  \includegraphics[width=8cm,draft=false]{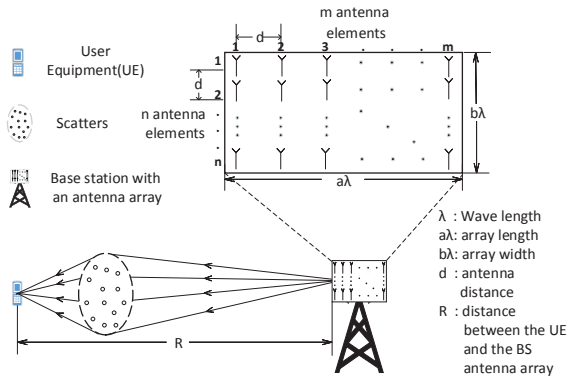}\\
  \caption{\small System model.}\label{fig1}
\end{figure}

A massive MIMO wireless transmission system is illustrated in Fig.~\ref{fig1}. The wireless down-link between a user equipment (UE) with multi-antenna and a BS with a 2D rectangular antenna array is studied in this paper.

First of all, we define some basic parameters for this model. We define $\lambda $ as the wavelength of the carrier, $d$ as the antenna spacing of this antenna array, $a\lambda {\rm{ }} \left({a \ge 1}\right) $ as the length of this antenna array and $b\lambda {\rm{ }}\left( {b \ge 1} \right)$ as the width of this antenna array. If we would like to deploy $m$ antennas in each row and $n$ antennas in each column for this antenna array, then we will have the relationship as listed in (1),
\[d = \frac{{a\lambda }}{{m - 1}} = \frac{{b\lambda }}{{n - 1}},\tag{1}\]
and the total number of antennas in this antenna array $M$ can be derived easily as
\[M = mn.\tag{2}\]

If we define $SN{R_{BS}}$ as the signal-to-noise ratio (SNR) at the BS, ${\mathbf{H}}$ and $\beta$ stand for the small scale fading matrix and large scale fading coefficient of the channel in this model respectively, the signal the BS transmits is defined as ${\mathbf{x}}$, $\mathbf{w}$ means the additive white Gaussian noise (AWGN) over wireless channels, and the mutual coupling matrix is configured as ${\mathbf{K}}$, equivalent precoding matrix is configured as ${{\mathbf{F}}_{eq}}$, ${\mathbf{A}}$ is defined as steering matrix, then the down-link signal vector received at a UE equipped with $N$ antennas can be expressed as
\[{\mathbf{y}} = \sqrt {SN{R_{BS}}} {\mathbf{HAK}}{{\mathbf{F}}_{eq}}{\beta ^{1/2}}{\mathbf{x}} + {\mathbf{w}},\tag{3}\]
in which ${\mathbf{x}}$ is a ${{N_s} \times 1}$ vector, $\mathbf{w}$ is a ${{N} \times 1}$ vector.

${\mathbf{H}} \sim \mathcal{C}\mathcal{N}\left( {0,{P{\mathbf{I}}}} \right)$ is governed by a complex Gaussian distribution, and is expressed as
\[{\mathbf{H}} = {\left[ {{\mathbf{h}_1},...,{\mathbf{h}_p},...,{\mathbf{h}_P}} \right]^T} \in {\mathbb{C}^{N \times P}} ,\tag{4}\]
in which ${\mathbb{C}^{N \times P}}$ denotes a $N \times P$ matrix, $P$ stands for the number of the independent incident directions, ${\mathbf{h}_p} \sim \mathcal{C}\mathcal{N}\left( {0,{{\mathbf{I}}}} \right)$ stands for the complex coefficient vector of small scale fading received from the $p$th incident direction, which is expressed as
\[{\mathbf{h}_p} = \mathbf{h}_p^{(r)} + j\mathbf{h}_p^{(i)} ,\tag{5}\]
in which $\mathbf{h}_p^{(r)}$ is defined as the real part of ${\mathbf{h}_p}$, and $\mathbf{h}_p^{(i)}$ is defined as the imaginary part of ${\mathbf{h}_p}$. Furthermore, both of them are Gaussian random variables distributed independently and identically, whose expectation and variance are $0$ and $0.5$ respectively.

Definitely, $P$ will be very large if considerable scatterers exist in the propagation environment. According to \cite{Ngo13, H.Q.11}, we divide the angular domain into $P$ independent incident directions with $P$ being large but finite.

Here we assume both of the azimuth angle ${\phi _q}$($q = 1,...,P$) and elevation angle $\theta$ are within the scope of $\left[ {{{ - \pi } \mathord{\left/
 {\vphantom {{ - \pi } 2}} \right.
 \kern-\nulldelimiterspace} 2},{\pi  \mathord{\left/
 {\vphantom {\pi  2}} \right.
 \kern-\nulldelimiterspace} 2}} \right]$. Each independent incident direction corresponds to one steering vector ${\mathbf{a}}\left( {{\phi _q},\theta } \right) \in {\mathbb{C}^{M \times 1}}$, so all the $P$ steering vectors can constitute the steering matrix ${\bf{A}}$ of the rectangular antenna array which is expressed as
\[{\mathbf{A}} = \left[ {{\mathbf{a}}\left( {{\phi _1},\theta } \right),...,{\mathbf{a}}\left( {{\phi _q},\theta } \right),...,{\mathbf{a}}\left( {{\phi _P},\theta } \right)} \right].\tag{6}\]
If we define ${{\mathbf{A}}^q} \in {\mathbb{C}^{n \times m}}$ as the steering matrix of the $q$th incident direction of the rectangular antenna array, we will get the following relationship,
\[{\mathop{\rm vec}\nolimits} \left( {{{\bf{A}}^q}} \right) = {\bf{a}}\left( {{\phi _q},\theta } \right),\tag{7}\]
in which $\operatorname{vec} \left(  \cdot  \right)$ is defined as the matrix vectorization operation.

Without loss of generality, we assume the antenna which locates at the first place for both the row and the column of the rectangular antenna array as the reference point of which the phase response is zero. And we normalize amplitude responses of all the antennas of the antenna array as 1. We define ${\bf{A}}_{ce}^q$, $\left( {1 \le c \le m,{\rm{ }}1 \le e \le n} \right)$ as the element in the steering matrix ${{\bf{A}}^q}$ which locates at the $c$th row and $e$th column, and it is expressed as
\[{\mathbf{A}}_{ce}^q = \exp \left\{ {j\frac{{2\pi }}{\lambda }\left[ \begin{gathered}
  {\text{(}}c - 1{\text{)}}dcos{\phi _q}sin\theta  \hfill \\
   + (e - 1)dsin{\phi _q}sin\theta  \hfill \\
\end{gathered}  \right]} \right\}.\tag{8}\]

For a rectangular antenna array with M elements, we define ${\mathbf{{\bf K}}} \in {\mathbb{C}^{M \times M}}$ as the corresponding mutual coupling matrix, which is expressed as \cite{B.Clerckx07}
\[{\bf{{\bf K}}} = {Z_L}{\left( {{Z_L}{{\bf{I}}} + {{\bf{Z}}_{{M}}}} \right)^{ - 1}},\tag{9}\]
in which ${Z_L}$ denotes the antenna load impedance that is constant for each antenna, ${{\bf{Z}}_{{M}}}$ denotes the $M \times M$ mutual impedance matrix, and ${{\bf{I}}}$ denotes an $M \times M$ unit matrix. From Fig. 1, ${{\bf{Z}}_{{M}}}$ can be constructed by $n \times n$ sub-matrices, i.e.,${{\bf{Z}}_{{M}}} = {\left[ {{{\bf{Z}}_{st}}} \right]_{n \times n}}$, where ${{\bf{Z}}_{st}}$, as an $m \times m$ mutual impedance sub-matrix, denotes the mutual impedances between the $m$ antennas located at the $s$th $(s = 1,...,n)$ row and the $m$ antennas located at the $t$th $(t = 1,...,n)$ row in the rectangular antenna array. For ease of exposition, we define $An{t_{su}}$ as the antenna located at the $s$th row and $u$th $(s = 1,...,m; u = 1,...,m)$ column of the rectangular antenna array, and define $An{t_{tv}}$ as the antenna located at the $t$th row and the $v$th $(t = 1,...,m; v = 1,...,m)$ column of the rectangular antenna array, the corresponding distance between which is given as $d_{uv}^{st} = d\sqrt {{{(t - s)}^2} + {{(v - u)}^2}}$. Thus ${{\bf{Z}}_{st}}$ can be written as
 \[{{\bf{Z}}_{{{st}}}} = \left[ {\begin{array}{*{20}{c}}
{z_{11}^{st}}&{z_{12}^{st}}&{...}&{z_{1m}^{st}}\\
{z_{21}^{st}}&{z_{22}^{st}}&{...}&{z_{2m}^{st}}\\
 \vdots & \vdots & \ddots & \vdots \\
{z_{m1}^{st}}&{z_{m2}^{st}}&{...}&{z_{mm}^{st}}
\end{array}} \right],\tag{10}\]

Consider a special case where all $M$ antenna elements in the rectangular antenna array are dipole antennas with the same parameters. Then the mutual impedance $z_{uv}^{st}$ only depends on the antenna spacing and can be obtained with the EMF method in \cite{Balanis12}. With a fixed antenna spacing $d$, we have the following properties:
\[z_{uv}^{st} = z_{\left( {u + 1} \right),\left( {v + 1} \right)}^{st},\tag{11}\]
\[z_{uv}^{st} = z_{vu}^{st}.{\kern 1pt} {\kern 1pt} {\kern 1pt} {\kern 1pt} {\kern 1pt} {\kern 1pt} {\kern 1pt} {\kern 1pt} {\kern 1pt} {\kern 1pt} {\kern 1pt} {\kern 1pt} {\kern 1pt} {\kern 1pt} {\kern 1pt} {\kern 1pt} {\kern 1pt} {\kern 1pt} {\kern 1pt} {\kern 1pt} {\kern 1pt} {\kern 1pt} {\kern 1pt} {\kern 1pt} {\kern 1pt} {\kern 1pt} {\kern 1pt} {\kern 1pt} {\kern 1pt} {\kern 1pt} \tag{12}\]
Similar properties can be derived for ${{\bf{Z}}_{st}}$ as
\[{{\mathbf{Z}}_{st}} = {{\mathbf{Z}}_{\left( {s + 1} \right),\left( {t + 1} \right)}},\tag{13}\]
\[{{\mathbf{Z}}_{st}} = {{\mathbf{Z}}_{ts}}.{\kern 1pt} {\kern 1pt} {\kern 1pt} {\kern 1pt} {\kern 1pt} {\kern 1pt} {\kern 1pt} {\kern 1pt} {\kern 1pt} {\kern 1pt} {\kern 1pt} {\kern 1pt} {\kern 1pt} {\kern 1pt} {\kern 1pt} {\kern 1pt} {\kern 1pt} {\kern 1pt} {\kern 1pt} {\kern 1pt} {\kern 1pt} {\kern 1pt} {\kern 1pt} {\kern 1pt} {\kern 1pt} {\kern 1pt} {\kern 1pt} {\kern 1pt} {\kern 1pt} {\kern 1pt} {\kern 1pt} {\kern 1pt} \tag{14} \]

Together with (10)-(14), the mutual impedance matrix ${{\bf{Z}}_{{M}}}$ can be readily obtained. It bears noting that with (10)-(14), the computational complexity can be significantly reduced compared to the direct calculation of the $M \times M$ entries of ${{\bf{Z}}_{{M}}}$, especially with a large $M$.

The equivalent precoding matrix ${{\mathbf{F}}_{{\text{eq}}}} = {{\mathbf{F}}_{{\text{RF}}}}{{\mathbf{F}}_{{\text{BB}}}}$ consists of baseband precoding matrix ${{\mathbf{F}}_{BB}}$ and the RF precoding matrix${{\mathbf{F}}_{RF}}$.

${N_s}$ data streams are transmitted by $N_{RF}^t$ radio frequency (RF) chains and $M$ antennas at the BS. All wireless data is received by $N_{RF}^r$ RF chains and $N$ antennas at the UE. In this case, the detected wireless signals at the UE is expressed by
\[ \mathbf{\tilde y} = {\mathbf{W}}_{eq}^\dag \mathbf{y} = {\mathbf{W}}_{BB}^\dag {\mathbf{W}}_{RF}^\dag \mathbf{y}, \tag{15}\]
in which $\dag$ is a conjugate transpose operation, ${\mathbf{W}}_{eq}$ is a $N \times {N_s}$ equivalent signal detection matrix which consists of baseband detection matrix ${{\mathbf{W}}_{BB}}$ and the RF detection matrix ${{\mathbf{W}}_{RF}}$, $\mathbf{y}$ is the received signal vector at antennas of the UE. Essentially, ${{\mathbf{F}}_{RF}}$ and ${{\mathbf{W}}_{RF}}$ are phase shift matrices used for the signal precoding and detection at the RF chains. Hence, the absolute value of the RF detection matrix ${{\mathbf{F}}_{RF}}$ and the RF precoding matrix ${{\mathbf{F}}_{RF}}$ is equal to 1.

\section{Mutual Coupling Effect Modeling}

\begin{figure*}[!t]
\[{R_{\max }} = \log \left| {{\mathbf{I}} + \frac{{SN{R_{BS}}}}{{{N_s}}}{\mathbf{W}}_{eq}^\dag {\mathbf{HAK}}{{\mathbf{F}}_{eq}}{\mathbf{F}}_{eq}^\dag {{\mathbf{K}}^{\mathbf{\dag }}}{{\mathbf{{A}}}^{\mathbf{\dag }}}{{\mathbf{H}}^{\mathbf{\dag }}}{{\mathbf{W}}_{eq}}{{\mathbf{R}}^{ - 1}}} \right|, \tag{21}\]
with \[{\mathbf{R}} = {\mathbf{W}}_{eq}^\dag {{\mathbf{W}}_{eq}}.\tag{21.1}\]
\end{figure*}

\subsection{Receive Diversity Gain Models}

Deployed in a constrained space at the BS, the number of antenna elements is inversely proportional to the antenna spacing, i.e., a larger number of antennas lead to a smaller antenna spacing. As concluded in \cite{C.Masouros13}, more antennas lead to a higher receive diversity gain of the massive MIMO system, whereas the diversity gain can be compromised by the mutual coupling effect that is caused by decreasing the antenna spacing. Thus, when a number of antennas are deployed in a fixed constrained area, there exists a tradeoff between $M$ and $d$, and it is important to analyze the the effect of mutual coupling on the achievable receive diversity gain of the massive MIMO systems.

Firstly, with a fixed antenna spacing, the antenna number receive diversity gain ${\mathbb{G}_M}$ is defined as
\[{\mathbb{G}_M} = \xi _M^{{d_{\min }}} - \xi _{{M_{{\text{min}}}}}^{{d_{\min }}},\tag{16}\]
in which $\xi _M^{{d_{\min }}}$ denotes the expectation of the receive SNR at the UE with $M$ antennas and minimum antenna spacing being $d_{\min }$ at the antenna array of the BS, and $\xi _{{M_{{\text{min}}}}}^{{d_{\min }}}$ denotes the expectation of the receive SNR at the UE subject to the minimum antenna spacing $d_{\min }$ and minimum antenna number ${M_{\min }}$ at the antenna array of the BS.

Secondly, with a fixed number of antennas, the antenna spacing receive diversity gain ${\mathbb{G}_d}$ is defined as
\[{\mathbb{G}_d} = \xi _{{M_{\min }}}^d - \xi _{{M_{\min }}}^{{d_{\min }}},\tag{17}\]
in which $\xi _{{M_{\min }}}^d$ denotes the expectation of the receive SNR at the UE with an antenna spacing of $d$ and with at least ${M_{\min }}$ antennas at the antenna array of the BS. $\xi _{{M_{{\text{min}}}}}^{{d_{\min }}}$, which can be considered as the baseline for $\mathbb{G}_d$ and $\mathbb{G}_M$, is the same as that in (16).

In order to obtain the expectation of the received SNR in (16) and (17), perfect channel state information is assumed to be available at the BS, which uses maximal-ratio combining (MRC) for signal detection. When there are $M$ antennas and the neighboring antennas are separated with a spacing of $d$, the received signal after MRC detection at the UE is given as \cite{Tse05}
\[\tilde y = {{\bf{G}}^\dag }{\bf{y}}{\kern 1pt}  = \sqrt {SN{R_{UE}}} {{\bf{G}}^\dag }{\bf{G}}x{\kern 1pt}  + {{\bf{G}}^\dag }{\bf{w}},\tag{18a}\]
in which ${{\mathbf{G}}^\dag }$ denotes the conjugate transpose of ${\mathbf{G}}$ with
\[{\mathbf{G}} = {\mathbf{HA{K}}}{{\mathbf{F}}_{eq}}{\beta ^{1/2}}.\tag{18b}\]

In addition, the average SNR seen at the UE side can be written as
\[\begin{gathered}
  SN{R_{UE}} = \frac{{SN{R_{BS}}{{\left\| {{{\mathbf{G}}^{\mathbf{\dag }}}{\mathbf{G}}} \right\|}^2}}}{{\left\| {{{\mathbf{G}}^{\mathbf{\dag }}}{\mathbf{G}}} \right\|}} \hfill \\
  {\kern 1pt} {\kern 1pt} {\kern 1pt} {\kern 1pt} {\kern 1pt} {\kern 1pt} {\kern 1pt} {\kern 1pt} {\kern 1pt} {\kern 1pt} {\kern 1pt} {\kern 1pt} {\kern 1pt} {\kern 1pt} {\kern 1pt} {\kern 1pt} {\kern 1pt} {\kern 1pt} {\kern 1pt} {\kern 1pt} {\kern 1pt} {\kern 1pt} {\kern 1pt} {\kern 1pt} {\kern 1pt} {\kern 1pt} {\kern 1pt} {\kern 1pt} {\kern 1pt} {\kern 1pt} {\kern 1pt} {\kern 1pt} {\kern 1pt}  = SN{R_{BS}}\left\| {{{\mathbf{G}}^{\mathbf{\dag }}}{\mathbf{G}}} \right\| \hfill \\
\end{gathered} ,\tag{19}\]

Then with $M$ antenna elements with antenna spacing $d$, the expectation of the SNR at the UE can be obtained as
\[\begin{gathered}
  \xi _M^d = {E}\left\{ {SN{R_{UE}}} \right\} \hfill \\
  {\kern 1pt} {\kern 1pt} {\kern 1pt} {\kern 1pt} {\kern 1pt} {\kern 1pt} {\kern 1pt} {\kern 1pt} {\kern 1pt} {\kern 1pt} {\kern 1pt} {\kern 1pt} {\kern 1pt}  = SN{R_{BS}}\mathbb{E}\left\{ {\left\| {{{\mathbf{G}}^\dag }{\mathbf{G}}} \right\|} \right\} \hfill \\
  {\kern 1pt} {\kern 1pt} {\kern 1pt} {\kern 1pt} {\kern 1pt} {\kern 1pt} {\kern 1pt} {\kern 1pt} {\kern 1pt} {\kern 1pt} {\kern 1pt} {\kern 1pt} {\kern 1pt}  = SN{R_{BS}}N\beta \left\| {{\mathbf{F}}_{eq}^\dag {{\mathbf{K}}^{\mathbf{\dag }}}{{\mathbf{{A}}}^{\mathbf{\dag }}}{{\mathbf{H}}^{\mathbf{\dag }}}{\mathbf{HA{K}}}{{\mathbf{F}}_{eq}}} \right\| \hfill \\
\end{gathered},\tag{20} \]
in which $E\left\{  \cdot  \right\}$ denotes the expectation operation. We can further obtain ${\mathbb{G}_d}$ and ${\mathbb{G}_M}$ through substituting (20) into (16) and (17), and replacing $M$ and $d$ with $M_{\min }$ and $d_{\min }$.

\subsection{Shannon Capacity with Optimal RF Chains}

\begin{figure*}[!t]
\[\begin{gathered}
  {R_{\max }} = \log \left| {{\mathbf{I}} + \frac{{SN{R_{BS}}}}{{{N_s}}}{\mathbf{HF}}{{\mathbf{F}}^\dag }{{\mathbf{H}}^\dag }{{\mathbf{W}}_{eq}}{{\left( {{\mathbf{W}}_{eq}^\dag {{\mathbf{W}}_{eq}}} \right)}^{ - 1}}{\mathbf{W}}_{eq}^\dag } \right| \hfill \\
   {\kern 1pt} {\kern 1pt} {\kern 1pt} {\kern 1pt} {\kern 1pt} {\kern 1pt} {\kern 1pt} {\kern 1pt} {\kern 1pt} {\kern 1pt} {\kern 1pt} {\kern 1pt} {\kern 1pt} {\kern 1pt} {\kern 1pt} {\kern 1pt} {\kern 1pt} {\kern 1pt} {\kern 1pt} {\kern 1pt} = \log \left| {{\mathbf{I}} + \frac{{SN{R_{BS}}}}{{{N_s}}}{{\mathbf{U}}_{\mathbf{H}}}{{\mathbf{\Sigma }}_{\mathbf{H}}}{\mathbf{V}}_{\mathbf{H}}^\dag {{\mathbf{U}}_{\mathbf{F}}}{{\mathbf{\Sigma }}_{\mathbf{F}}}{\mathbf{\Sigma }}_{\mathbf{F}}^\dag {\mathbf{U}}_{\mathbf{F}}^\dag {{\mathbf{V}}_{\mathbf{H}}}{\mathbf{\Sigma }}_{\mathbf{H}}^\dag {\mathbf{U}}_{\mathbf{H}}^\dag } \right| \hfill \\
   {\kern 1pt} {\kern 1pt} {\kern 1pt} {\kern 1pt} {\kern 1pt} {\kern 1pt} {\kern 1pt} {\kern 1pt} {\kern 1pt} {\kern 1pt} {\kern 1pt} {\kern 1pt} {\kern 1pt} {\kern 1pt} {\kern 1pt} {\kern 1pt} {\kern 1pt} {\kern 1pt} {\kern 1pt} {\kern 1pt} = \log \left| {{\mathbf{I}} + \frac{{SN{R_{BS}}}}{{{N_s}}}{\mathbf{V}}_{\mathbf{H}}^\dag {{\mathbf{U}}_{\mathbf{F}}}{{\mathbf{\Sigma }}_{\mathbf{F}}}{\mathbf{\Sigma }}_{\mathbf{F}}^\dag {\mathbf{U}}_{\mathbf{F}}^\dag {{\mathbf{V}}_{\mathbf{H}}}{\mathbf{\Sigma }}_{\mathbf{H}}^\dag {{\mathbf{\Sigma }}_{\mathbf{H}}}} \right| \hfill \\
\end{gathered} ,\tag{25}\]

\end{figure*}

A phase shift matrix is designed to separate the RF chains and the antennas. Assume that the relationship among the numbers of data stream, RF chains and antennas is configured as ${N_s} \leqslant N_{RF}^t \leqslant M$. Considering the equivalent precoding matrix ${\left( {{{\mathbf{F}}_{eq}}} \right)_{M \times {N_s}}} = {\left( {{{\mathbf{F}}_{RF}}} \right)_{M \times N_{RF}^t}}{\left( {{{\mathbf{F}}_{BB}}} \right)_{N_{RF}^t \times {N_s}}}$, about it's rank, we have $rank\left( {{{\mathbf{F}}_{eq}}} \right) \leqslant \min \left( {{N_s},N_{RF}^t,M} \right)$. This result implies that the up-bound of the degree of freedom at the equivalent precoding matrix is depended on the minimum among the numbers of data stream, RF chains and antennas. When the number of RF chains is larger than the number of data stream, a part of number of RF chains, i.e., ${N_{RF}^t}-{N_s}$, has not been utilized by the equivalent precoding matrix. To save the cost, the number of RF chains can be configured as ${{N_s}}$ to satisfy the requirement of the equivalent precoding matrix ${{\mathbf{F}}_{eq}}$.

Based on the system model in Fig. 1, the system achievable rate, i.e., the maximum Shannon capacity is expressed by (21) and normalized on the unit bandwidth \cite{zhang05}.
Let $MIN = \min \left( {M,P} \right)$, the eigenvalues of wireless channels ${\mathbf{H}}$ are ordered by ${\lambda _1} \geqslant {\lambda _2} \cdots  \geqslant {\lambda _{MIN}}$. The maximum available rate is ${R_{\max }} = \sum\limits_{k = 1}^{MIN} {\log \left( {1 + \frac{{SN{R_{BS}}}}{{{N_s}}}{\lambda _k}} \right)} $, where ${SN{R_{BS}}}$ is the SNR value at the BS. The rank of wireless channel ${\mathbf{H}}$ is denoted by $r = rank\left( {\mathbf{H}} \right)$. As a consequence, the maximum available rate is rewritten by ${R_{\max \_full}} = \sum\limits_{k = 1}^{rank\left( {\mathbf{H}} \right)} {\log \left( {1 + \frac{{SNR\_BS}}{{{N_s}}}{\lambda _k}} \right)} $. When the numbers of data stream and RF chains are equal to the rank of wireless channel $rank\left( {\mathbf{H}} \right)$, the wireless channel capacity has been fully utilized. When the rank of wireless channels $rank\left( {\mathbf{H}} \right)$ is less than the number of data stream, what should we do to configure the number of RF chains? To utilize the wireless channel capacity and save the implementation cost of RF chains, the number of RF chains is configured as the rank of wireless channels in this paper.

To simplify the derivation, the rank of wireless channels is assumed to be larger than the number of data stream in the following study. The optimal equivalent detection matrix ${{\mathbf{W}}_{eq}}$ is derived by a singular value decomposition (SVD) method
\[{{\mathbf{W}}_{eq}} = {{\mathbf{U}}_{\mathbf{W}}}{{\mathbf{\Sigma }}_{\mathbf{W}}}{\mathbf{V}}_{\mathbf{W}}^\dag ,\tag{22}\] with \[{{\mathbf{\Sigma }}_{\mathbf{W}}} = \left( \begin{gathered}
  {{\mathbf{\Delta }}_{\mathbf{W}}} \hfill \\
  {\mathbf{0}} \hfill \\
\end{gathered}  \right), \tag{22.1}\]
\[{{\mathbf{\Delta }}_{\mathbf{W}}} = \left( {\begin{array}{*{20}{c}}
  {{w_1}}&0& \cdots &0 \\
  0&{{w_2}}& \cdots &0 \\
   \vdots & \vdots & \ddots & \vdots  \\
  0&0& \cdots &{{w_r}}
\end{array}} \right),\tag{22.2}\]
in which ${{\mathbf{U}}_{\mathbf{W}}}$ and ${\mathbf{V}}_{\mathbf{W}}^\dag$ are unitary matrices.

The optimal equivalent precoding matrix ${{\mathbf{F}}_{eq}}$ is derived by a SVD method
\[{\mathbf{F}}_{eq} = {{\mathbf{U}}_{\mathbf{F}}}{{\mathbf{\Sigma }}_{\mathbf{F}}}{\mathbf{V}}_{\mathbf{F}}^\dag ,\tag{23}\] with \[{{\mathbf{\Sigma }}_{\mathbf{F}}} = \left( {\begin{array}{*{20}{c}}
  {{{\mathbf{\Delta }}_{\mathbf{F}}}}&{\mathbf{0}} \\
  {\mathbf{0}}&{\mathbf{0}}
\end{array}} \right),\tag{23.1}\] \[{{\mathbf{\Delta }}_{\mathbf{F}}} = \left( {\begin{array}{*{20}{c}}
  {{f_1}}&0& \cdots &0 \\
  0&{{f_2}}& \cdots &0 \\
   \vdots & \vdots & \ddots & \vdots  \\
  0&0& \cdots &{{f_r}}
\end{array}} \right),\tag{23.2}\]
in which ${{\mathbf{U}}_{\mathbf{F}}}$ and ${\mathbf{V}}_{\mathbf{F}}^\dag$ are unitary matrices.

\begin{figure*}[!t]

\[\begin{gathered}
  {R_{\max }} = \log \left| {\frac{{SN{R_{BS}}}}{{{N_s}}}\left( {\frac{{{N_s}}}{{SN{R_{BS}}}}{\mathbf{I}} + {{\mathbf{\Sigma }}_{{{\mathbf{F}}^2}}}{{\mathbf{\Sigma }}_{{{\mathbf{H}}^2}}}} \right)} \right| \hfill \\
  {\kern 1pt} {\kern 1pt} {\kern 1pt} {\kern 1pt} {\kern 1pt} {\kern 1pt} {\kern 1pt} {\kern 1pt} {\kern 1pt} {\kern 1pt} {\kern 1pt}  {\kern 1pt} {\kern 1pt} {\kern 1pt} {\kern 1pt} {\kern 1pt} {\kern 1pt} {\kern 1pt} {\kern 1pt} {\kern 1pt} {\kern 1pt} = r\log \left( {\frac{{SN{R_{BS}}}}{{{N_s}}}} \right) + \log \left( {\mathop \Pi \limits_{i = 1}^r \left( {\frac{{{N_s}}}{{SN{R_{BS}}}} + f_i^2\lambda _i^2} \right)} \right) \hfill \\
\end{gathered},\tag{32} \]

\[\begin{gathered}
  L\left( {f_1^2,f_2^2 \cdots f_r^2} \right) \hfill \\
   = \left( {\frac{{{N_s}}}{{SN{R_{BS}}}} + \lambda _1^2f_1^2} \right)\left( {\frac{{{N_s}}}{{SN{R_{BS}}}} + \lambda _2^2f_2^2} \right) \cdots \left( {\frac{{{N_s}}}{{SN{R_{BS}}}} + \lambda _r^2f_r^2} \right) \hfill \\
   + \alpha \left( {f_1^2 + f_2^2 +  \cdots  + f_r^2 - N_s^2} \right) \hfill \\
\end{gathered} .\tag{33}\]

\[\begin{gathered}
  \frac{{\partial L\left( {f_1^2,f_2^2 \cdots f_r^2} \right)}}{{\partial f_i^2}} \hfill \\
  \;\;\;\; = \lambda _i^2\left( {\frac{{{N_s}}}{{SN{R_{BS}}}} + \lambda _1^2f_1^2} \right) \cdots \left( {\frac{{{N_s}}}{{SN{R_{BS}}}} + \lambda _{i - 1}^2f_{i - 1}^2} \right) \hfill \\
  \;\;\;\;\;\;\; \times \left( {\frac{{{N_s}}}{{SN{R_{BS}}}} + \lambda _{i + 1}^2f_{i + 1}^2} \right) \cdots \left( {\frac{{{N_s}}}{{SN{R_{BS}}}} + \lambda _r^2f_r^2} \right) + \alpha  \hfill \\
\end{gathered} ,\tag{34}\]
\end{figure*}

When a SVD method is performed over the equivalent channel ${{\mathbf{H}}_{{\text{eq}}}} = {\mathbf{HAK}}$, the equivalent channel is derived by
\[{{\mathbf{H}}_{{\text{eq}}}} = {{\mathbf{U}}_{\mathbf{H}}}{{\mathbf{\Sigma }}_{\mathbf{H}}}{\mathbf{V}}_{\mathbf{H}}^\dag ,\tag{24}\] with \[{{\mathbf{\Sigma }}_{\mathbf{H}}} = \left( {\begin{array}{*{20}{c}}
  {{{\mathbf{\Delta }}_{\mathbf{H}}}}&{\mathbf{0}} \\
  {\mathbf{0}}&{\mathbf{0}}
\end{array}} \right),\tag{24.1}\] \[{{\mathbf{\Delta }}_{\mathbf{H}}} = \left( {\begin{array}{*{20}{c}}
  {{\lambda _1}}&0& \cdots &0 \\
  0&{{\lambda _2}}& \cdots &0 \\
   \vdots & \vdots & \ddots & \vdots  \\
  0&0& \cdots &{{\lambda _r}}
\end{array}} \right),\tag{24.2}\]
in which ${{\mathbf{U}}_{\mathbf{H}}}$ and ${\mathbf{V}}_{\mathbf{H}}^\dag$ are unitary matrices.

Based on (22), (23) and (24), the maximum available rate ${R_{\max }}$ is further derived by a SVD method in (25) when the optimal equivalent detection matrix ${{\mathbf{W}}_{eq}}$ is assumed to be a non-singular matrix. Let
\[{{\mathbf{\Sigma }}_{{{\mathbf{F}}^2}}} = {{\mathbf{\Sigma }}_{\mathbf{F}}}{\mathbf{\Sigma }}_{\mathbf{F}}^\dag  = \left( {\begin{array}{*{20}{c}}
  {{{\mathbf{\Delta }}_{\mathbf{F}}}{\mathbf{\Delta }}_{\mathbf{F}}^\dag }&{\mathbf{0}} \\
  {\mathbf{0}}&{\mathbf{0}}
\end{array}} \right),\tag{26}\]

\[{{\mathbf{\Sigma }}_{{{\mathbf{H}}^2}}} = {\mathbf{\Sigma }}_{\mathbf{H}}^\dag {{\mathbf{\Sigma }}_{\mathbf{H}}} = \left( {\begin{array}{*{20}{c}}
  {{\mathbf{\Delta }}_{\mathbf{H}}^\dag {{\mathbf{\Delta }}_{\mathbf{H}}}}&{\mathbf{0}} \\
  {\mathbf{0}}&{\mathbf{0}}
\end{array}} \right),\tag{27}\]

 \[{\mathbf{U}} = {\mathbf{V}}_{\mathbf{H}}^\dag {{\mathbf{U}}_{\mathbf{F}}}.\tag{28}\]
(25) is rewritten by
\[\begin{gathered}
  R = \log \left| {{\mathbf{I}} + \frac{{SN{R_{BS}}}}{{{N_s}}}{\mathbf{U}}{{\mathbf{\Sigma }}_{{{\mathbf{F}}^2}}}{{\mathbf{U}}^\dag }{{\mathbf{\Sigma }}_{{{\mathbf{H}}^2}}}} \right| \hfill \\
\end{gathered} ,\tag{29}\]
in which ${{\mathbf{\Sigma }}_{{{\mathbf{F}}^2}}}$ and ${{\mathbf{\Sigma }}_{{{\mathbf{H}}^2}}}$ are diagonal matrices and the values of elements at diagonal line are larger than zero. Furthermore, the eigenvalues of ${{\mathbf{\Sigma }}_{{{\mathbf{F}}^2}}}$ and ${{\mathbf{\Sigma }}_{{{\mathbf{H}}^2}}}$ are the elements at the diagonal lines, respectively. $ {\mathbf{U}}$ is a unitary matrix, i.e., ${\left\| {\mathbf{U}} \right\|_2} = M$. When $ {\mathbf{U}}$ is configured as a diagonal matrix, the maximum available rate is achieved by
\[\begin{gathered}
R_{\max} = \log \left| {\mathbf{I} + \frac{{SN{R_{BS}}}}{{{N_s}}}{{\mathbf{\Sigma }}_{{{\mathbf{F}}^2}}}{{\mathbf{\Sigma }}_{{{\mathbf{H}}^2}}}} \right| \hfill \\
\end{gathered} .\tag{30}\]
Assume that the transmission power at the BSs is independent of the equivalent precoding matrix. This assumption implies that ${\left\| {{{\mathbf{F}}_{eq}}} \right\|_F}{\text{ = }}{N_s}$, i.e.,
\[\sum\limits_{i = 1}^r {f_i^2}  = N_s^2 .\tag{31}\]
The maximum available rate $R_{max}$ can be simplified as (32).

To achieve the maximum achievable rate, the optimal solution of the equivalent precoding is derived by a Lagrange multiplier method in the following. A function is first constructed in (33) with Lagrange factor $\alpha$. And then take the derivative of $L\left( {f_1^2,f_2^2 \cdots f_r^2} \right)$ with respect to $f_i^2,\left( {i = 1,2...r} \right)$ in (34).
\begin{figure*}[!t]
\[{{\mathbf{F}}_{eq}} = \underbrace {\left[ {\begin{array}{*{20}{c}}
  {{e^{j{\vartheta _{1,1}}}}}& \cdots &{{e^{j{\vartheta _{1,2r}}}}}&0& \cdots &0 \\
   \vdots & \cdots & \vdots & \vdots & \cdots & \vdots  \\
  {{e^{j{\vartheta _{M,1}}}}}& \cdots &{{e^{j{\vartheta _{M,2r}}}}}&0& \cdots &0
\end{array}} \right]}_{{{\left( {{{\mathbf{F}}_{RF}}} \right)}_{M \times 2{N_s}}}}\underbrace {\left[ {\begin{array}{*{20}{c}}
  {{b_1}}&0& \cdots &0&0&{\begin{array}{*{20}{c}}
   \cdots &0
\end{array}} \\
  {{b_1}}&0& \cdots &0&0&{\begin{array}{*{20}{c}}
   \cdots &0
\end{array}} \\
  0&{{b_2}}& \cdots &0&0&{\begin{array}{*{20}{c}}
   \cdots &0
\end{array}} \\
  0&{{b_2}}& \cdots &0&0&{\begin{array}{*{20}{c}}
   \cdots &0
\end{array}} \\
   \vdots & \vdots & \ddots & \vdots & \vdots &{\begin{array}{*{20}{c}}
   \cdots & \vdots
\end{array}} \\
  0&0& \cdots &{{b_r}}&0&{\begin{array}{*{20}{c}}
   \cdots &0
\end{array}} \\
  0&0& \cdots &{{b_r}}&0&{\begin{array}{*{20}{c}}
   \cdots &0
\end{array}} \\
  0&0& \cdots &0&0&{\begin{array}{*{20}{c}}
   \cdots &0
\end{array}} \\
   \vdots & \vdots & \ddots & \vdots & \vdots &{\begin{array}{*{20}{c}}
   \ddots & \vdots
\end{array}} \\
  0&0& \cdots &0&0&{\begin{array}{*{20}{c}}
   \cdots &0
\end{array}}
\end{array}} \right]}_{{{\left( {{{\mathbf{F}}_{BB}}} \right)}_{2{N_s} \times {N_s}}}}, \tag{40}\]

\[\begin{gathered}
  {C_E}\left( \theta  \right) =  - \frac{1}{{\theta T}}\ln \left( {E\left\{ {{e^{ - \theta TB\left( {\log \left( {\frac{{SN{R_{BS}}}}{{{N_s}}}} \right) + \log \left( {\mathop \Pi \limits_{i = 1}^r \left( {\frac{{{N_s}}}{{SN{R_{BS}}}} + f_i^2\lambda _i^2} \right)} \right)} \right)}}} \right\}} \right) \hfill \\
  {\kern 1pt} {\kern 1pt} {\kern 1pt} {\kern 1pt} {\kern 1pt} {\kern 1pt} {\kern 1pt} {\kern 1pt} {\kern 1pt} {\kern 1pt} {\kern 1pt} {\kern 1pt} {\kern 1pt} {\kern 1pt} {\kern 1pt} {\kern 1pt} {\kern 1pt} {\kern 1pt} {\kern 1pt} {\kern 1pt} {\kern 1pt} {\kern 1pt} {\kern 1pt} {\kern 1pt} {\kern 1pt} {\kern 1pt} =  - \frac{1}{{\theta T}}\ln \left( {{e^{ - \theta TB\log \left( {\frac{{SN{R_{BS}}}}{{{N_s}}}} \right)}}E\left\{ {{{\left( {{e^{\log \left( {\mathop \Pi \limits_{i = 1}^r \left( {\frac{{{N_s}}}{{SN{R_{BS}}}} + f_i^2\lambda _i^2} \right)} \right)}}} \right)}^{ - \theta TB}}} \right\}} \right) \hfill \\
  {\kern 1pt} {\kern 1pt} {\kern 1pt} {\kern 1pt} {\kern 1pt} {\kern 1pt} {\kern 1pt} {\kern 1pt} {\kern 1pt} {\kern 1pt} {\kern 1pt} {\kern 1pt} {\kern 1pt} {\kern 1pt} {\kern 1pt} {\kern 1pt} {\kern 1pt} {\kern 1pt} {\kern 1pt} {\kern 1pt} {\kern 1pt} {\kern 1pt} {\kern 1pt} {\kern 3pt} =  - \frac{1}{{\theta T}}\ln \left( {{e^{ - \theta TB\log \left( {\frac{{SN{R_{BS}}}}{{{N_s}}}} \right)}}E\left\{ {{{\left( {\mathop \Pi \limits_{i = 1}^r \left( {\frac{{{N_s}}}{{SN{R_{BS}}}} + f_i^2\lambda _i^2} \right)} \right)}^{ - \frac{{\theta TB}}{{\ln \left( 2 \right)}}}}} \right\}} \right) \hfill \\
\end{gathered}.\tag{42} \]

\end{figure*}

Let \[\frac{{\partial L\left( {f_1^2,f_2^2 \cdots f_r^2} \right)}}{{\partial f_i^2}} = 0 ,\tag{35}\]
in which $\left( {i = 1,2...r} \right)$, we can further derive the following result\[\begin{gathered}
  \frac{{\frac{{{N_s}}}{{SN{R_{BS}}}}}}{{\lambda _i^2}} + f_i^2 = \frac{{\frac{{{N_s}}}{{SN{R_{BS}}}}}}{{\lambda _j^2}} + f_j^2 \hfill \\
   \Rightarrow {\kern 1pt} {\kern 1pt} {\kern 1pt} {\kern 1pt} {\kern 1pt} {\kern 1pt} {\kern 1pt} {\kern 1pt} {\kern 1pt} {\kern 1pt} f_i^2 - f_j^2 = \frac{{{N_s}}}{{SN{R_{BS}}}}\left( {\frac{1}{{\lambda _j^2}} - \frac{1}{{\lambda _i^2}}} \right) \hfill \\
\end{gathered} ,\tag{36}\]
in which $\left( {i,j = 1,2...r} \right)$.

Based on (31), the square of eigenvalues at the equivalent precoding matrix ${{\mathbf{F}}_{eq}}$ is derived by
\[\begin{gathered}
  f_i^2 = \frac{{N_s}{\left( {\frac{1}{{\lambda _1^2}} + \frac{1}{{\lambda _2^2}} +  \cdots  + \frac{1}{{\lambda _r^2}}} \right)}}{{SN{R_{BS}}r}} \hfill \\
  {\kern 1pt} {\kern 1pt} {\kern 1pt} {\kern 1pt} {\kern 1pt} {\kern 1pt} {\kern 1pt} {\kern 1pt} {\kern 1pt} {\kern 1pt} {\kern 1pt} + \frac{{N_s^2}}{r} - \frac{{{N_s}}}{{SN{R_{BS}}\lambda _i^2}} \hfill \\
\end{gathered} .\tag{37}\]

According to (25) and (30), we know that ${{\mathbf{W}}_{eq}}$, ${{\mathbf{V}}_{\mathbf{F}}}$, ${{\mathbf{U}}_{\mathbf{H}}}$ and ${\mathbf{U}}$ are removed in the simplification process of ${R_{\max }}$, so ${{\mathbf{V}}_{\mathbf{F}}}$, ${{\mathbf{U}}_{\mathbf{H}}}$ and ${\mathbf{U}}$ can be unit matrices. And furthermore, ${{\mathbf{U}}_{\mathbf{F}}}$ and ${{\mathbf{V}}_{\mathbf{H}}}$ can also be unit matrices, and ${{\mathbf{U}}_{\mathbf{F}}} = {{\mathbf{V}}_{\mathbf{H}}}$. Considering the ${\mathbf{W}}_{eq}$ is a $N \times {N_s}$  non-singular matrix, ${\mathbf{W}}_{eq}$ can be decomposed by
\[{{\mathbf{W}}_{eq}} = \left( {\begin{array}{*{20}{c}}
  {\mathbf{I}} \\
  {\mathbf{0}}
\end{array}} \right) ,\tag{38}\]
In this case, the optimal equivalent precoding matrix is simplified by
\[\begin{gathered}
  {{\mathbf{F}}_{eq}} = {{\mathbf{U}}_{\mathbf{F}}}{{\mathbf{\Sigma }}_{\mathbf{F}}}{\mathbf{V}}_{\mathbf{F}}^\dag  \hfill \\
  {\kern 1pt} {\kern 1pt} {\kern 1pt} {\kern 1pt} {\kern 1pt} {\kern 1pt} {\kern 1pt} {\kern 1pt} {\kern 1pt} {\kern 1pt} {\kern 1pt} {\kern 1pt} {\kern 1pt} {\kern 1pt} {\kern 1pt} {\kern 1pt}  = {{\mathbf{V}}_{\mathbf{H}}}{{\mathbf{\Sigma }}_{\mathbf{F}}} \hfill \\
\end{gathered} .\tag{39}\]

Based on the method in \cite{zhang05}, the optimal equivalent precoding matrix ${{\mathbf{F}}_{eq}}$ is composed of $ {{\mathbf{F}}_{RF}}$ and ${{\mathbf{F}}_{BB}}$, which are designed in (40) with

\[{b_j} = \frac{1}{2}\mathop {\max }\limits_{1 \leqslant i \leqslant M} \left| {{f_{i,j}}} \right|, \tag{40.1}\]
\[{\vartheta _{i,\left( {2j - 1} \right)}} = \measuredangle {f_{i,j}} - {\cos ^{ - 1}}\frac{{\left| {{f_{i,j}}} \right|}}{{2{b_j}}} ,\tag{40.2}\]
\[{\kern 11pt} {\kern 1pt} {\kern 3pt} {\vartheta _{i,2j}} = \measuredangle {f_{i,j}} + {\cos ^{ - 1}}\frac{{\left| {{f_{i,j}}} \right|}}{{2{b_j}}},\tag{40.3}\]
in which ${{f_{i,j}}}$ is the element of ${{\mathbf{F}}_{eq}}$ located at the $i$th row and the $j$th column，$\measuredangle {f_{i,j}}$ is the corresponding angle.

Based on the result in (40), the number of RF chains $2{N_s}$ can satisfy the requirement of the optimal equivalent precoding matrix. In general, the number of antennas $M$ is larger than the number of RF chains $2{N_s}$ in 5G massive MIMO wireless systems. Hence, our proposed optimal equivalent precoding matrix can save the number of RF chains $M - 2{N_s}$.

\subsection{Effective Capacity with Mutual Coupling Effect}

From \cite{Xiaohu14}, we define the effective capacity under multimedia constraints as
\[{C_E}\left( \theta  \right) = - \frac{1}{{\theta T}}\ln \left( {E\{ {e^{ - \theta TBR}}\} } \right), \tag{41}\]
in which $\theta $ and B denote the QoS statistical exponent and bandwidth respectively, ${\emph{E}\{\cdot\}}$ is the expectation operation. Without losing generality, we consider independent fading channels that keep static within a frame duration $T$.

\begin{figure*}[!t]
\[\begin{gathered}
  E\left\{ {{R_{\max }}} \right\} = E\left\{ {r\log \left( {\frac{{SN{R_{BS}}}}{{{N_s}}}} \right) + \log \left( {\mathop \Pi \limits_{i = 1}^r \left( {\frac{{{N_s}}}{{SN{R_{BS}}}} + f_i^2\lambda _i^2} \right)} \right)} \right\} \hfill \\
  {\kern 1pt} {\kern 1pt} {\kern 1pt} {\kern 1pt} {\kern 1pt} {\kern 1pt} {\kern 1pt} {\kern 1pt} {\kern 1pt} {\kern 1pt} {\kern 1pt} {\kern 1pt} {\kern 1pt} {\kern 1pt} {\kern 1pt} {\kern 1pt} {\kern 1pt} {\kern 1pt} {\kern 1pt} {\kern 1pt} {\kern 1pt} {\kern 1pt} {\kern 1pt} {\kern 1pt} {\kern 1pt} {\kern 1pt} {\kern 1pt} {\kern 1pt} {\kern 1pt} {\kern 1pt} {\kern 1pt} {\kern 1pt} {\kern 1pt} {\kern 1pt} {\kern 1pt} {\kern 1pt} {\kern 1pt} {\kern 1pt} {\kern 1pt} {\kern 1pt} {\kern 1pt} {\kern 1pt}  \leqslant {r\log \left( {\frac{{SN{R_{BS}}}}{{{N_s}}}} \right) + \sum\limits_{i = 1}^r {\log \left( {E\left\{ {\frac{{{N_s}}}{{SN{R_{BS}}}} + f_i^2\lambda _i^2} \right\}} \right)} } \hfill \\
\end{gathered},\tag{44} \]

\[\begin{gathered}
  E\{ {R_{\max }}\}  \leqslant  {r\log \left( {\frac{{SN{R_{BS}}}}{{{N_s}}}} \right) + \sum\limits_{i = 1}^r {\log \left( {\frac{{{N_s}}}{r}E\left\{ {\left( {\lambda _i^2{N_s} + \lambda _i^2\frac{{\sum\limits_{i = 1}^r {\frac{1}{{\lambda _i^2}}} }}{{SN{R_{BS}}}}} \right)} \right\}} \right)} } \hfill \\
  {\kern 1pt} {\kern 1pt} {\kern 1pt} {\kern 1pt} {\kern 1pt} {\kern 1pt} {\kern 1pt} {\kern 1pt} {\kern 1pt} {\kern 1pt} {\kern 1pt} {\kern 1pt} {\kern 1pt} {\kern 1pt} {\kern 1pt} {\kern 1pt} {\kern 1pt} {\kern 1pt} {\kern 1pt} {\kern 1pt} {\kern 1pt} {\kern 1pt} {\kern 1pt} {\kern 1pt} {\kern 1pt} {\kern 1pt} {\kern 1pt} {\kern 1pt} {\kern 1pt} {\kern 1pt} {\kern 1pt} {\kern 1pt} {\kern 1pt} {\kern 1pt} {\kern 1pt} {\kern 1pt} {\kern 1pt} {\kern 1pt}  \leqslant  {r\log \left( {\frac{{SN{R_{BS}}}}{{{N_s}}}} \right) + \sum\limits_{i = 1}^r {\log \left( {\frac{{{N_s}}}{{{r^2}}}\left( {{N_s} + \frac{r}{{SN{R_{BS}}}}} \right)E\left\{ {\sum\limits_{i = 1}^r {\lambda _i^2} } \right\}} \right)} }  \hfill \\
  {\kern 1pt} {\kern 1pt} {\kern 1pt} {\kern 1pt} {\kern 1pt} {\kern 1pt} {\kern 1pt} {\kern 1pt} {\kern 1pt} {\kern 1pt} {\kern 1pt} {\kern 1pt} {\kern 1pt} {\kern 1pt} {\kern 1pt} {\kern 1pt} {\kern 1pt} {\kern 1pt} {\kern 1pt} {\kern 1pt} {\kern 1pt} {\kern 1pt} {\kern 1pt} {\kern 1pt} {\kern 1pt} {\kern 1pt} {\kern 1pt} {\kern 1pt} {\kern 1pt} {\kern 1pt} {\kern 1pt} {\kern 1pt} {\kern 1pt} {\kern 1pt} {\kern 1pt}  =  {r\log \left( {\frac{{SN{R_{BS}}}}{{{N_s}}}} \right) + \sum\limits_{i = 1}^r {\log \left( {\frac{{{N_s}}}{{{r^2}}}\left( {{N_s} + \frac{r}{{SN{R_{BS}}}}} \right)E\left\{ {tr\left( {{{\mathbf{H}}^\dag }{\mathbf{HAK}}{{\mathbf{K}}^\dag }{{\mathbf{A}}^\dag }} \right)} \right\}} \right)} }  \hfill \\
\end{gathered} ,\tag{45}\]

\[\begin{gathered}
  E\{ {R_{\max }}\}  \leqslant  {r\log \left( {\frac{{SN{R_{BS}}}}{{{N_s}}}} \right) + r\log \left( {\frac{{{N_s}}}{{{r^2}}}\left( {{N_s} + \frac{r}{{SN{R_{BS}}}}} \right)\left( {tr\left( {{\mathbf{AK}}{{\mathbf{K}}^\dag }{{\mathbf{A}}^\dag }} \right) + E\left\{ {tr\left( {{{\mathbf{H}}^\dag }{\mathbf{H}}} \right)} \right\}} \right)} \right)} \hfill \\
  {\kern 1pt} {\kern 1pt} {\kern 1pt} {\kern 1pt} {\kern 1pt} {\kern 1pt} {\kern 1pt} {\kern 1pt} {\kern 1pt} {\kern 1pt} {\kern 1pt} {\kern 1pt} {\kern 1pt} {\kern 1pt} {\kern 1pt} {\kern 1pt} {\kern 1pt} {\kern 1pt} {\kern 1pt} {\kern 1pt} {\kern 1pt} {\kern 1pt} {\kern 1pt} {\kern 1pt} {\kern 1pt} {\kern 1pt} {\kern 1pt} {\kern 1pt} {\kern 1pt} {\kern 1pt} {\kern 1pt} {\kern 1pt} {\kern 1pt} {\kern 1pt} {\kern 1pt} {\kern 1pt} {\kern 1pt} {\kern 1pt} {\kern 1pt}  =  {r\log \left( {\frac{{SN{R_{BS}}}}{{{N_s}}}} \right) + r\log \left( {\frac{{{N_s}}}{{{r^2}}}\left( {{N_s} + \frac{r}{{SN{R_{BS}}}}} \right)\left( {tr\left( {{\mathbf{AK}}{{\mathbf{K}}^\dag }{{\mathbf{A}}^\dag }} \right) + \Pr } \right)} \right)}  \hfill \\
\end{gathered} ,\tag{46}\]

\[{C_E}\left( \theta  \right) \leqslant B\left( {r\log \left( {\frac{{SN{R_{BS}}}}{{{N_s}}}} \right) + r\log \left( {\frac{{{N_s}}}{{{r^2}}}\left( {{N_s} + \frac{r}{{SN{R_{BS}}}}} \right)\left( {tr\left( {{\mathbf{AK}}{{\mathbf{K}}^\dag }{{\mathbf{A}}^\dag }} \right) + \Pr } \right)} \right)} \right).\tag{47}\]

\end{figure*}

Considering the maximum available rate in (32), (41) can be extended as (42), and it is clear that $f\left( x \right) = {x^{ - a}},\left( {a > 0} \right)$ is a convex function. Then an upper bound of the effective capacity can be obtained using Jensen's inequality,
\[\begin{gathered}
  {C_E}\left( \theta  \right) =  - \frac{1}{{\theta T}}\ln \left( {E\left\{ {{e^{ - \theta TBR}}} \right\}} \right) \hfill \\
   {\kern 27pt}\leqslant  - \frac{1}{{\theta T}}\ln \left( {{e^{ - \theta TBE\left\{ R \right\}}}} \right) \hfill \\
   {\kern 27pt}= BE\left\{ R \right\} \hfill \\
\end{gathered}.\tag{43}\]
Based on (32), the upper bound of the maximum available rate is derived in (44). When the optimal equivalent precoding matrix is used for massive MIMO wireless systems, the upper bound of the maximum available rate is further derived in (45). Considering the lemma 2.9 in \cite{Tulino04}, the upper bound of the maximum available rate is finally expressed in (46). As a consequence, the upper bound of the effective capacity in 5G multimedia communication systems is given by (47).

\section{Numerical Results and Analysis}

In this section, we demonstrate the performance of the multimedia oriented massive MIMO communication systems in terms of the receive diversity gain as well as the effective capacity, where both effects of the QoS statistical exponent and mutual coupling are evaluated. For ease of illustration, we consider a rectangular antenna array with the length-width ratio of ${a \mathord{\left/
 {\vphantom {a b}} \right.
 \kern-\nulldelimiterspace} b} = 2$. There are altogether 128 dipole antenna elements in the rectangular antenna array \cite{J.Hoydis11}, each with length and diameter of $0.5\lambda $ and $0.001\lambda $, respectively. Then, a reasonable minimum antenna spacing is $d_{min} = 0.1\lambda$. The large scale fading factor is normalized to $\beta  = 1$ \cite{H.Q.11,C.Masouros13,Ngo13another}, with a load impedance at each antenna as ${Z_L} = 50$ Ohms \cite{Shen10}. Without loss of generality, we assume that the BS is located in rich scattering environment where the incident directions can arrive at an arbitrary angle uniformly. Thus  it is reasonable to assume that the elevation angle $\theta $ and azimuth ${\phi _q}$ follow {\em i.i.d.} uniform distributions within $\left[ { - \pi /2,\pi /2} \right]$. For ease of demonstration, the default number of independent incident directions $P = 70$ is configured, with frame duration $T = 1$ms and bandwidth $B = 1$MHz \cite{Xiaohu14}.

\begin{figure}
  \centering
  \includegraphics[width=7.7cm,draft=false]{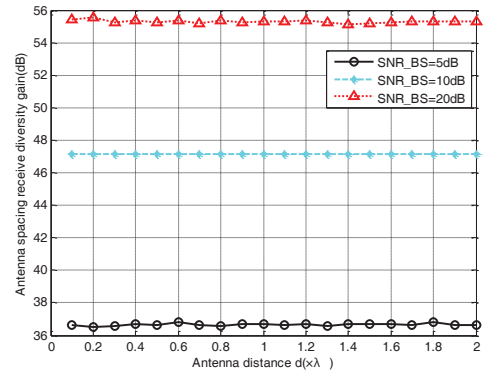}\\
  \caption{\small Antenna spacing receive diversity gain with respect to the antenna spacing considering different SNRs.}
  \label{fig2}
\end{figure}

In Fig.~\ref{fig2}, the antenna spacing receive diversity gain ${\mathbb{G}_d}$ with respect to the antenna spacing is investigated. ${M_{\min }}$ is configured as $1$. When the antenna spacing $d$ is fixed, the antenna spacing receive diversity gain increases with the increase of SNR. But if we fix the SNR at the BS, it is shown that there is almost no correlation between the antenna spacing receive diversity gain and the antenna spacing.

\begin{figure}
  \centering
  \includegraphics[width=7.7cm,draft=false]{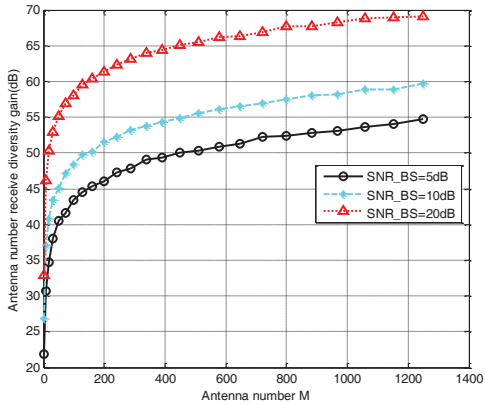}\\
  \caption{\small Antenna number receive diversity gain with respect to the antenna number considering different SNRs.}
  \label{fig3}
\end{figure}

Fig.~\ref{fig3} illustrates the correlation between ${\mathbb{G}_M}$ and the antenna number considering different SNRs. $d_{\min}$ is configured as $0.1\lambda$. And it is shown that the antenna number receive diversity gain has a positive correlation with the antenna number and SNR.

\begin{figure}
  \centering
  \includegraphics[width=8.5cm,draft=false]{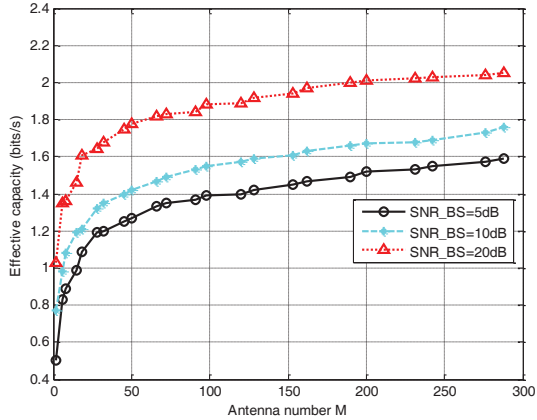}\\
  \caption{\small Effective capacity with respect to the antenna number considering different SNRs.}
  \label{fig4}
\end{figure}

In Fig.~\ref{fig4}, the effective capacity is illustrated with varying values of the antenna number and SNR. For ease of illustration, the antenna spacing is set as 0.5$\lambda$ and $\theta$ is set as 0.01. With a fixed SNR value, it is observed that a higher effective capacity is obtained by increasing the antenna number. In addition, with a fixed number of antennas, a higher effective capacity is obtained with a higher SNR.

\begin{figure}
  \centering
  \includegraphics[width=8.5cm,draft=false]{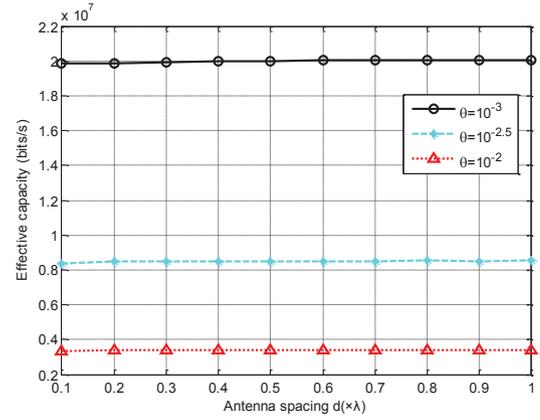}\\
  \caption{\small Effective capacity with respect to the QoS statistical exponent $\theta$ and the antenna spacing.}
  \label{fig5}
\end{figure}

Fig.~\ref{fig5} shows the effective capacity with varying values of QoS statistical exponent and the antenna spacing. With a fixed antenna spacing, it is observed that a higher effective capacity is reached by decreasing the QoS statistical exponent. On the other hand, for a fixed QoS statistical exponent and increasing of antenna spacing, the effective capacity almost keeps stationary.

\begin{figure}
  \centering
  \includegraphics[width=8.5cm,draft=false]{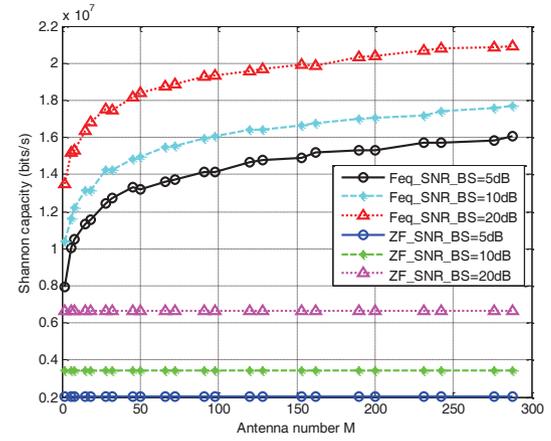}\\
  \caption{\small Shannon Capacity with respect to the antenna number considering different precoding matrices.}
  \label{fig6}
\end{figure}

When the number of user and the baseband data stream are configured as one, Fig.~\ref{fig6} compares the Shannon capacity with respect to the antenna number considering different precoding matrices. the proposed optimal equivalent precoding matrix labeled as ``Feq" and the zero-forcing precoding matrix labeled as ``ZF" are compared in Fig.~\ref{fig6}. When the number of antennas and the SNR are fixed, the Shannon capacities with the proposed optimal equivalent precoding matrix are greater than the Shannon capacities with the zero-forcing precoding matrix. Moreover, the Shannon capacities with the proposed optimal equivalent precoding matrix has a positive correlation with the the number of antennas. However, the Shannon capacities with the zero-forcing precoding matrix almost keeps stationary with the increase of the number of antennas. This result indicates that our proposed optimal equivalent precoding matrix can improve the Shannon capacity, i.e., the available rate in massive MIMO wireless communication systems.

\begin{figure}
  \centering
  \includegraphics[width=8.5cm,draft=false]{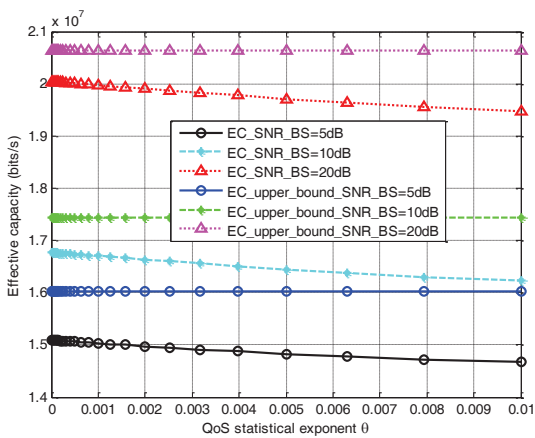}\\
  \caption{\small Effective capacity with respect to the QoS statistical exponent $\theta$ considering different SNRs.}
  \label{fig7}
\end{figure}
Fig.~\ref{fig7} analyzes the effective capacity and the upper bound of effective capacity with respect to the QoS statistical exponent considering different SNRs, in which ``EC\_SNR" labels the effective capacity results and ``EC\_upper\_bound\_SNR" represents the upper bound of the effective capacity results. When the SNR is fixed, there is a positive correlation between the effective capacity and the QoS statistical exponent. Moreover, the more the the QoS statistical exponent decrease, the closer the upper bound of effective capacity gets to the effective capacity.

\begin{figure}
  \centering
  \includegraphics[width=8.5cm,draft=false]{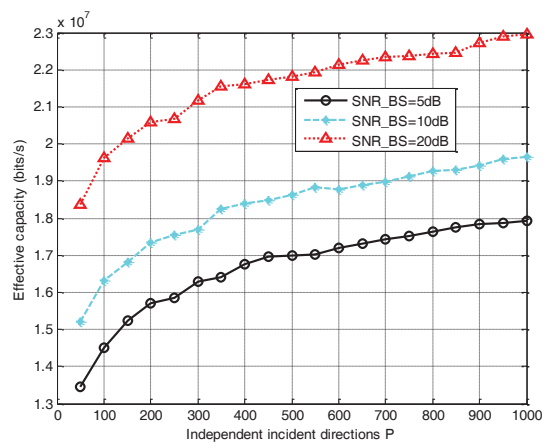}\\
  \caption{\small Effective capacity with respect to the independent incident directions $P$ considering different SNRs.}
  \label{fig8}
\end{figure}
When the number of antennas is configured as 128, Fig.~\ref{fig8} describes the effective capacity with respect to the number of independent incident directions. When the SNR is fixed, there is a positive correlation between the effective capacity and the independent incident directions.

\section{Conclusions}

Based on the mutual coupling effect, an optimal equivalent precoding matrix has been proposed to maximize the available rate and save the cost of RF chains for 5G massive MIMO multimedia communication systems. Considering the requirements of multimedia wireless communications, the upper bound of the effective capacity has been derived for 5G massive MIMO multimedia communication systems with the QoS statistical exponent constraint. Compared with the conventional ZF precoding matrix, numerical results show that the proposed optimal equivalent precoding matrix can obviously improve the available rate for 5G massive MIMO multimedia communication systems. In the future work, taking into account the QoS statistical exponent constraints, a more efficient signal detection precoding algorithm is worth exploring towards better performance of the multimedia massive MIMO communication systems.

\acks The authors would like to acknowledge the support from the National Natural Science Foundation of China (NSFC) under the grants 60872007, 61301128, 61461136004 and 61271224, NFSC Major International Joint Research Project under the grant 61210002, the Ministry of Science and Technology (MOST) of China under the grants 2015FDG12580 and 2014DFA11640, the Fundamental Research Funds for the Central Universities under the grant 2015XJGH011 and 2013ZZGH009. This research is partially supported by the EU FP7-PEOPLE-IRSES, project acronym S2EuNet (grant no. 247083), project acronym WiNDOW (grant no. 318992) and project acronym CROWN (grant no. 610524).

\end{document}